\documentclass[12pt,prd,showpacs,amsmath,amssymb,aps,floats]{revtex4-1}%
\usepackage{epsfig}
\usepackage{dcolumn}
\usepackage{bm}
\usepackage{amsmath}
\usepackage{amsfonts}
\usepackage{amssymb}
\def\nnd{\end{document}}

\def\be{\begin{equation}}
\def\ee{\end{equation}}

\newcommand{\bea}{\begin{eqnarray}}
\newcommand{\eea}{\end{eqnarray}}
\newcommand{\bwt}{\begin{widetext}}
\newcommand{\ewt}{\end{widetext}}

\def\u
\def\hZ{\widehat Z}

\def\eed{\end{document}}

\def\m_z{m_{\textrm {Z}}}

\renewcommand{\u}{\rm{u}}

\def\be{\beta}

\def\sl#1{#1\!\!\!\!/}
\def\rm#1{\textrm{#1}}

\makeatother
\begin{document}
\title{The $b'$ search at the LHC}
\author{B.~Holdom$^1$}
\author{Qi-Shu Yan$^{1,2}$}
\affiliation{$^1$ Department of Physics, University of Toronto, Toronto, Canada }
\affiliation{$^2$ College of Physics Sciences, Graduate University of Chinese Academy of Sciences, \\ Beijing 100039, P.R China }

\begin{abstract}
We consider the production and detection of a sequential, down type quark via the mode $p p \to b' {\bar b}' \to W^+ W^- t {\bar t} \to \ell \nu_{\ell} 8 j$ at the LHC, with the collision energy $\sqrt{s}=10$ TeV and the total integrated luminosity around $1$ fb$^{-1}$. We assume $m_{b'}=m_{t'}=600$ GeV. A full reconstruction is employed and the signal and background discrimination is studied within a neural network approach. Our results show that this mode can make a useful contribution to the $b'$ search.
\end{abstract}

\maketitle

\renewcommand{\thefootnote}{\arabic{footnote}} \setcounter{footnote}{0}%


A sequential fourth family is a well-defined extension of the standard model \cite{Frampton:1999xi,Holdom:2009rf} which could have important implications for our understanding of electroweak symmetry breaking and the flavor problem \cite{Holdom:2006mr,Holdom:2010za}. In order to confirm or rule out the fourth family at the LHC it will be important to be able to test for the existence of both the $t'$ and $b'$. 
The detection of $t'$ could first occur in the lepton plus jets mode, as with the discovery of top quark, via $t' {\bar t}' \to W^+ W^- b {\bar b}$ where one $W$ decays leptonically \cite{Arik:1996qd}. The process $b' {\bar b}' \to W^+ W^- q\overline{q}$ is similar if it is assumed that $q$ is $u$ or $c$ \cite{Ozcan:2008qk}. 
But when $b'$ dominantly decays to $W^- t$ then the lepton plus jets mode faces a considerable combinatoric problem \cite{Holdom:2009rf,Burdman:2008qh,AguilarSaavedra:2009es}. 

The search for the $t'$ can benefit from the use of jet invariant masses to identify $W$-jets \cite{Holdom:2007nw,Holdom:2007ap,Holdom:2010za}, since the $W$'s are more likely to be both boosted and isolated when coming from the $t'\bar{t}'$ rather than from $t\bar{t}$. Here a relatively large cone size is used in the jet finding algorithm, and the $W$-jet can be simply combined with another jet to reconstruct the $t'$ mass. This is not as successful for the $b'$  since here the $W$ must be combined with a reconstructed $t$. This would require higher mass $b'$'s (as for the vectorlike quarks in \cite{Skiba:2007fw}) where the $t$'s are more boosted.

When two $W$'s decay leptonically then the same-sign lepton signal becomes available \cite{Contino:2008hi}. This has a branching fraction almost 6 times smaller than the single lepton mode, but it enjoys the advantage that the standard model background is small. The CMS collaboration has adopted this channel to explore the discovery  potential of $b' {\bar b}'$ \cite{cms}. However without reconstructing all objects in the decay chains it is difficult to distinguish a fourth family signature from that of other new physics producing a same-sign lepton signature. For instance this signature might come from fourth family neutrinos, especially when they have Majorana masses \cite{Holdom:2006mr,CuhadarDonszelmann:2008jp,Rajaraman:2010wk}. Same-sign $W$'s or $t$'s or charginos can appear in final states in SUSY models \cite{Kraml:2005kb,Alwall:2007ed}. A same-sign top pair can be produced by exchanging a neutral scalar or vector boson in $t$ and $u$ channels \cite{Larios:2003jq}. In flavor models for neutrino physics a doubly charged particle can give rise to same-sign leptons \cite{Chen:2008qb,delAguila:2008cj}. An attempt to reconstruct the other objects in the events containing same-sign leptons is hampered in the case of $b'\overline{b'}$ production in the context of a fourth family. In this case the same-sign leptons must emerge from different heavy quarks. In principle the transverse mass of each $b'$ can be reconstructed via the MT2 method \cite{Lester:1999tx} but one must again confront nontrivial combinatorics.

A comprehensive study of the heavy quarks in vectorlike models has been conducted by \cite{AguilarSaavedra:2009es}, where signatures with four-, three-, double-, and single-lepton + jets have been studied using the likelihood method. However, the study of the discovery potential for the lepton plus jets mode $b' {\bar b}' \to W^- t W^+ {\bar t} \to \ell \nu_{\ell} 8j$ is left undone apparently due to the large combinatorics. This mode was explored in the ATLAS Design Report \cite{atlas} where it also does not appear to be a useful discovery mode.  Here we shall explore whether these difficulties may be at least partially overcome by using a full reconstruction method combined with a neural network analysis. 

We are considering a complete sequential fourth family of chiral matter fields and we choose $m_{t'}=m_{b'}=600$ GeV. It is typically thought that $m_{t'}>m_{b'}$ from electroweak precision constraints \cite{Kribs:2007nz}. One of our goals is to separate the effects of the $t'$ from the $b'$ signal, and so a larger $t'$ would only make this easier. Thus in this sense our $m_{t'}$ choice is conservative. 600 GeV masses are close to the unitarity upper bound \cite{Chanowitz:1979}, and so this is also a conservative choice for considering the discovery reach. We assume that the $b'$ decays predominantly to $t$, as might be expected from an extended CKM matrx. It is certainly consistent with current bounds inferred from single production of $t$ at Tevatron \cite{Abazov:2009ii,Aaltonen:2009jj} and from the global fit from precision data and low energy processes \cite{Bobrowski:2009ng,Chanowitz:2009mz}. Assuming that branching fraction for $b' \to W t$ is unity, the decay modes and branching fractions from $b' {\bar b}'$ are as displayed in Table (\ref{table2}). 

\begin{table}[th]
\begin{center}%
\begin{tabular}
[c]{|c|c|c|c|c|c|}\hline
$b' {\bar b}'$ & $4\ell \sl{E} 2j$ & $3 \ell \sl{E}4j$ & $2 \ell \sl{E} 6j $ & $\ell  \sl{E} 8j$ & $10j$ \\\hline
BF$\times  100 (\ell=e,\mu,\tau$) & $1.0$ & $9.1$ & $28.5$ & $39.3$ & $20.4$ \\ \hline
BF$\times  100 \,\,\,\,\,\,(\ell=e,\mu$)   & $0.2$ & $2.4$ & $12.8$ & $26.2$  & $20.4$ \\\hline
\end{tabular}
\end{center}
\caption{Branching fractions of decay mode of $b' {\bar b}'$ are shown.}%
\label{table2}%
\end{table}

We shall attempt to reconstruct all objects in each event and to collectively use both the kinematic observables and the observables obtained from reconstructed objects to suppress background. Both heavy quarks in each event will be reconstructed. We will only consider the dominant backgrounds from $t {\bar t}+ nj$ and $W + nj$, since other backgrounds can be safely neglected \cite{Skiba:2007fw,atlas}. For the effects of QCD corrections we must estimate $K$ factors appropriate to the large center of mass energies and large number of jets in events passing our cuts. We use $K=1.5$ for both the signal events and the $t {\bar t} + nj$ background, and $K=1$ for the $W + nj$ background \cite{Cacciari:2008zb,Campbell:2003hd,Huston:2010}.

We use Madgraph/MadEvent \cite{Maltoni:2002qb} to generate signal events and Alpgen \cite{Mangano:2002ea} to generate background events. The MLM parton-jet matching method is used where for the $t {\bar t}+ nj$ samples, $p_{T\rm{min}}=100$ GeV, while for the $W + nj$ samples, $p_{T\rm{min}}=150$ GeV. These values for the Alpgen $p_{T\rm{min}}$ parameter are appropriate for the large center of mass energies . Alpgen is designed so that physical results are quite insensitive to the value of $P_{T\rm{min}}$ as long as extreme values are not taken. Changing $P_{T\rm{min}}$ amounts to a rebinning of the jet multiplicity samples, and our choice of $P_{T\rm{min}}$ implies that the background is dominated by the $t {\bar t}+ 1j$, $t {\bar t}+(\ge2j)$, $W + 2j$ and $W +(\ge3j)$ samples \cite{Holdom:2007ap}. Pythia \cite{Sjostrand:2006za} is used to simulate shower, fragmentation, hadronization and decay processes. PGS \cite{pgs} is used to simulate the detector effects and to find jets, leptons, and missing energy in each event. We modify the PGS code slightly so as to adopt the anti-$k_T$ algorithm \cite{Cacciari:2008gp} to find jets in the events. For the jet resolution parameter we choose $R=0.4$.

We adopt the following preselection rules:
\begin{itemize}
\item Jets are required to have $p_T(j)>20$ GeV.
\item There is only one energetic lepton with $p_T(\ell)>20$ GeV ($\ell=e\,,\,\,\mu$) and the missing energy satisfies $\sl{E}>20$ GeV. 
\item We impose $\hat{s}> 1200$ GeV and $H_T > 900$ GeV. $\hat{s}$ is the invariant mass obtained from the momentum sum over lepton, missing energy, and jets which pass all cuts. $H_T$ is the scalar sum of transverse momentum over lepton, missing energy, and jets which pass all cuts.
\end{itemize}
The selection efficiencies for each of these preselection rules can be found in Table (II). In Table (III) we show the jet multiplicity samples for $n_j=6$, $n_j=7$, and $n_j\geq 8$. This shows that the samples with $n_j=7$ and $n_j\geq 8$ have superior signal to background ratios and so we concentrate on those. We do not consider $b$ tagging since it does not help much to separate signal from the $t {\bar t}+ n j$ background.

\begin{table}[th]
\begin{center}%
\begin{tabular}
[c]{|c|c|c|c|c|}\hline
& $b' \bar{b'}$ & $t' \bar{t'}$& $W +  \textrm{jets}$ &  $t {\bar t} +\textrm{jets}$\\\hline
$p_T(\ell)>20\rm{GeV}$&$40\%$ & $29\%$ & $47\%$ & $25\%$\\ \hline
$\sl{E}>20\rm{GeV}$& $37\%$ & $28\%$ & $43\%$ & $24\%$ \\ \hline
$\hat{s}>1200\rm{GeV}\mbox{ \& } H_T>900\rm{GeV}$& $28\%$ & $22\%$& $18\%$ & $7\%$\\ \hline \hline
No. of Events with 1 fb$^{-1}$ & $66.8$ & $56.9$ & $831.1$ &
$536.1$ \\\hline
\end{tabular}
\end{center}
\caption{The selection efficiencies of the preselection rules are shown. Note that we have also used $H_T$ cuts (less than 900 GeV) in Alpgen to generate the background events. In the last row, we normalize the number of events by assuming the integrated luminosity as 1 fb$^{-1}$. $K$ factors are not included.}%
\label{table4}%
\end{table}

\begin{table}[th]
\begin{center}%
\begin{tabular}
[c]{|c|c|c|c|c|c|c|}\hline
& $b' \bar{b'}$ & $t' \bar{t'}$& $W + \textrm{jets}$&  $t {\bar t} + \textrm{jets}$ &$S/B$ &$S/\sqrt{S+B}$\\\hline
$n_j=6$ & $12.5$ & $9.4$ & $87.3$ & $125.8$ & $0.05$ & $0.80$\\\hline
$n_j=7$ & $ 16.5$ & $ 4.8$ & $40.7$ & $107.1$ & $0.11$& $1.26$\\ \hline
$n_j\geq 8$ & $26.7$ & $3.1$ & $22.8$ &
$124.3$ &0.18& $2.00$\\ \hline
\end{tabular}
\end{center}
\caption{The data samples of different jet multiplicities are shown. All events pass the preselection rules given in Table (II). $K$ factors are not included.}%
\label{table41}%
\end{table}

Our full reconstruction of four $W$'s, two $t$'s, and two $b'$'s is a follows.
\begin{itemize}
\item[1)] Find the two-fold solutions of the $z$-component of the neutrino momentum by assuming that the lepton and missing energy are from the $W$ leptonic decay.
\item[2)] Loop over all jets and combine 6 jets into 3 hadronic $W$'s. 
\item[3)] Pair four $W$ candidates (three hadronic and one leptonic $W$'s, two of which come from $t$) and two $b$ jet candidates into 2 $b^{\prime}$'s and evaluate the $\chi^2$ function, which is defined as
\bea
\chi^2 &=\sum_{i=1}^{2} \frac{|m_{W_{t_i}} - m_W^{PDG}|^2}{\sigma_{W_{t_i}}^2} + \sum_{i=1}^{2} \frac{|m_{W_i} - m_W^{PDG}|^2}{\sigma_W^2} \nonumber \\  & + \sum_{i=1}^2 \frac{|m_{t_i} - m_t^{PDG}|^2}{\sigma_t^2}+ \frac{|m_{b_1'}-m_{b_2'}|^2}{\sigma_{b'}^2}\,. \label{chi3}
\eea
$\sigma_{W_{t_i}}=11$ GeV, $\sigma_W=14$ GeV, $\sigma_t=20$ GeV, and $\sigma_{b'}=25$ GeV arise from the resolution of the calorimeters detectors where the $\sigma^2$'s are assumed to be the sum of those of the decay products.
\item[4)] For each event, from all possible pairings including the neutrino two-fold ambiguity, we choose the one with minimum $\chi^2$ as the right reconstruction of all objects. 
\end{itemize}
For the channel $n_j\geq 8$ we assume that all $W$ bosons have two jets while for the channel $n_j=7$ we assume that one of the hadronic $W$ bosons can be a single jet. 

To further enhance signal to background it appears that we must resort to multidimensional variable analysis methods, such as likelihood, boosted decision tree or neural network methods. We choose the last of these methods and use the multilayer perceptron method which has been implemented in the ROOT package.

The basic idea of this method in data analysis is to employ the high dimensional feature space to better separate signal and background. This method has a long history in particle physics \cite{Denby:1992hh}, it has been developed into quite a mature form \cite{Hocker:2007ht} and it has been widely used in top quark precision measurement \cite{Pleier:2008ig}. Multiple jet final states, for example the full hadronic $t{\bar t}$ events, have been successfully separated from the huge background by using the NN method \cite{:2007qf,Abazov:2006yb}, where data sample with $n_j=6,7,8$ are considered. It should thus be feasible to apply it to a $b'$ search at the LHC.

The discriminating observables can be divided into two groups, as observables obtained before and after the reconstruction procedure. The first group includes:
\begin{itemize}
\item the transverse momenta of the leading 4 jets, the lepton and the missing transverse energy
\item the leading 4 invariant masses of jets
\item $H_T$, $\hat{s}$, $\cal{A}$ (aplanarity), $\cal{S}$ spherity,
$\cal{C}$ (centrality), $p^{sum}_z$ (the scalar sum of $z$ component of momenta)
\item the number of jets with momentum larger than $120$ GeV, $60$ GeV, $30$ GeV, and $20$ GeV, respectively.
\item the first, second, and third minimum invariant mass $m(j_1,j_2)$ of pair of jets
\item the first, second, and third minimum $R(j_1,j_2)$ of pair of jets
\item the $\phi$ angle between lepton and missing energy
\end{itemize}
Most of these observables are adopted by the Tevatron groups for the $t$ measurements \cite{Abazov:2006yb,Abazov:2007kg,:2007qf}.

The second group of observables are:
\begin{itemize}
\item Masses and momenta of reconstructed objects, i.e. four $W$'s, two $t$'s, two $b$ jets, and two $b'$ heavy quarks.
\item The angle of the $b'$ relative to the $z$ direction in the center of mass frame of the event. By combining the charge of lepton with whether the leptonic $W$ is isolated (not from top) or nonisolated (from top),  we can infer which heavy quark is $b'$.
\item We also reconstruct all events in terms of a $t\bar{t}$ production hypothesis. Here the identification of objects is based on
\bea
\chi^2_{t{\bar t}} &=\sum_{i=1}^{2} \frac{|m_{W_{t_i}} - m_W^{PDG}|^2}{\sigma_{W_{t_i}}^2} + \sum_{i=1}^2 \frac{| m_{t_i} - m_t^{PDG}|^2}{\sigma_t^2}.
\eea
Then we use the angle between the two $t$'s and the angle between the $W$ and $b$ from each $t$ decay as discriminating observables. 
\end{itemize}

\begin{figure}[t]
\centerline{
\epsfxsize=16 cm \epsfysize=10 cm \epsfbox{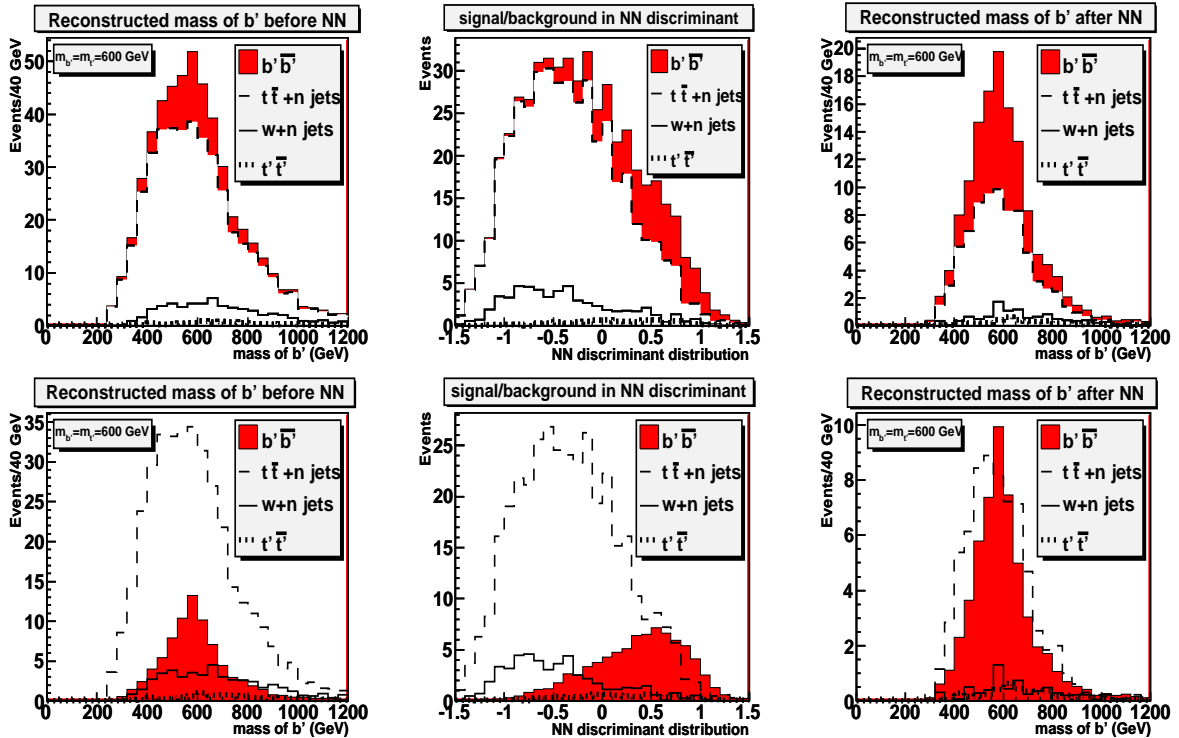}
}
\caption{The reconstructed mass peak of a 600 GeV mass $b'$ for both signal and background is shown for the $n_j \geq 8$ sample, where we provide both stacked (upper row) and unstacked (lower row) plots. The result before applying the neural network is shown in the left column and the neural network discriminant is shown in the middle column. With a discriminant cut of 0.15, the final result is shown in the right column. The $m_{b'}$ observable has been temporarily removed from the neural network to produce these plots. The assumed luminosity is 1 fb$^{-1}$and there are two contributions to the histogram from each event from the two values for $m_{b'}$. $K$ factors are included.}
\label{fig1}
\end{figure}

Our neural network has three layers: input layer, hidden layer, and output layer. For the $n_j \geq 8$ sample, we have ended up with 24 input layer and 48 hidden layer observables. For the $n_j=7$ sample, we have ended up with 18 input layer and 36 hidden layer observables. The output layer, valued from $-1$ to $1$, discriminates between background and signal. We use the default training method (BFGS) and the default parameters of the method encoded in the ROOT package. Other training methods in the ROOT package did not yield better results. We also find that various subsets of the observables we have chosen can yield quite similar results, but we did not find it worthwhile to try to minimize the number of observables.

We would like to compare the application of the NN method to the traditional counting method. For the latter we need to find a few most powerful discriminants which can separate the signal and background best and apply cuts to them sequentially, for instance, the reconstructed mass bumps of $b'$, $t$, and $W$, the $H_T$ distribution, the sphericity and $P_t$ of the leading 3 jets, etc. In our attempt to use this method we were not able to obtain significances much greater than unity. This shows how the NN method helps in the optimization of cuts. Furthermore, for our expanded set of kinematic observables the final performance of the NN method is quite stable and is almost independent of  the choice of variables. So in this sense the application of the NN method is more straightforward than the careful selection and tuning required in the traditional counting method. 

\begin{table}[t]
\begin{center}%
\begin{tabular}
[c]{|c|c|c|c|c|c|c|c|c|c|c|}\hline
& $b' \bar{b'}$ & $t' \bar{t'}$& $W + \textrm{jets}$ &  $t {\bar t} + \textrm{jets}$ & $\frac{S}{B}$ & $\frac{S}{\sqrt{B}}$ & $\frac{S}{\sqrt{S + B}}$ & $\frac{S}{\sqrt{B + (0.2B)^2}}$ & $\frac{S}{\sqrt{B + (0.1B)^2}}$ \\\hline \hline
$n_j=7$ & $11.3$ & $1.5$ & $8.0 $ &
$23.4  $ & $0.4 $ & $2.3 $ & $2.0 $ & $1.5$ & $2.0$\\ \hline
$n_j=7(n_b >0)$ & $9.0$ & $1.2$ & $0.8 $ &
$18.7  $ & $0.5 $ & $2.3  $ & $1.9$ & $1.7$ & $2.1$\\ \hline
\hline
$n_j\geq8$ & $18.7$ & $1.4$ & $3.7 $ &
$32.0  $  &  $0.6$& $3.4 $ & $2.7$ & $2.2$ & $2.9$ \\ \hline
$n_j\geq8 (n_b>0) $ & $14.9$& $ 1.1$ & $0.4 $ &
$25.6 $ &  $0.6$& $3.1  $ & $2.5$ & $2.2$ & $2.8$ \\ \hline
\end{tabular}
\end{center}
\caption{The number of events, normalized to 1 fb$^{-1}$, for both signal and background after applying neural network, for the $n_j=7$  and $n_j\geq 8$ jet multiplicities, and with and without a required $b$-tag. We combine the $t^{\prime} t^{\prime}$ and $b^{\prime} b^{\prime}$ events to define the signal $S$. The effects of the normalization uncertainty of the background processes are indicated in the last two columns. $K$ factors improve the results but are not included here (see the Summary).}%
\label{table5}%
\end{table}

In Fig. 1 we show the reconstructed mass peak of a $b'$ with a 600 GeV mass, with the integrated luminosity 1 fb$^{-1}$ from the $n_j \geq 8$ sample. The left and right columns show results without and with the neural network, while the middle shows the discriminant. We see that the $W+\rm{jets}$ background is small, which is largely the result of requiring a large number of jets. We find that the neural network discrimination improves $S/B$ by a factor of about $4$ for the $n_j=7$ sample, and $3$ for the $n_j\geq 8$ sample. And we also see that the $b' {\bar b}'$ signal can be quite effectively isolated from the contribution of $t' {\bar t}'$ with $m_{t'}=600$ GeV.

Table \ref{table5} summarizes our results. It also shows the effect of a modified reconstruction that makes use of at least one required $b$-tag. The result is a lower significance, which is not surprising since the dominant background from $t{\bar t}+\textrm{jets}$ also has $b$-jets. We used $b$-tagging efficiencies of 0.6, 0.1 and 0.01 for $b$, $c$ and light quarks respectively. There could be other reasons to employ $b$-tagging given that it reduces the combinatorics for proper event reconstruction. For example it could help to overcome the effects of pile-up in these high jet multiplicity events.

There is a systematic uncertainty on the overall normalization of the backgrounds due to our reliance on a Monte Carlo estimate. There are also other systematic uncertainties such as those related to the jet energy scale. To estimate the effects of such uncertainties, we show in the last two columns of Table \ref{table5} the significance in two cases where  the background normalization (for both $t{\bar t}$ and $W+\textrm{jets}$) is allowed to fluctuate up by $20\%$ and $10\%$ (a method used for example in \cite{Cheung:2011vx}). Statistical uncertainties will be of lesser importance as the integrated luminosity is increased.

\section*{Summary}
We have studied the specific case of  $m_{b'}=m_{t'}=600$ GeV with $\sqrt{s}=10$ TeV as a case study to test the feasibility of a full reconstruction method for the mode $pp\to b' {\bar b}' \to W^+ {\bar t} W^- t \to  \ell \nu_{\ell} 8 j$. This appears to be workable even though there are 7 or 8 jets and large combinatorics (over $10^4$). By applying the neural network technique the significance $S/\sqrt{S+B}$ can reach $4.0$. (From Table IV after combining two multiplicity samples we have $S/\sqrt{S+B}=3.3\times \sqrt{1.5} =4.0$ where $\sqrt{1.5}$ is from the $K$ factors.)  The significance drops to 2.6 or 3.8 for $20\%$ or $10\%$ upward fluctuations in the background. When a $b$-tag is required the significance is $3.8$. This is for an integrated luminosity $1$ fb$^{-1}$.  For $\sqrt{s}=7$ or 14 TeV the significance changes by a factor of about 1/2 and 3 respectively, and so in the case of 7 TeV this reduction will be overcome by a larger integrated luminosity.

To address the systematic uncertainties we expect that a more data-driven approach to background subtraction will be adopted by experimentalists. We would like to point out from Fig.~(1) that the discriminant distributions (middle plots) can be at least as useful as the mass distributions in the development of a subtraction procedure. Improvements in our analysis could come from a jet finding algorithm that is optimized for the treatment of highly boosted $W$'s that cannot be resolved as two jets. Information on the substructure of these massive jets may be useful or alternatively a continuous jet cone algorithm \cite{Borjanovic:2004ce} might be used, where the jet cone size can be smaller than $0.4$. 

We can compare the sensitivity of the lepton plus jets mode to the two lepton same-sign mode. The latter mode was studied in \cite{Holdom:2010za} for the same $\sqrt{s}$, luminosity and masses as considered here.  With the following restrictions on event selection,  $H_T>1$ TeV,  2 isolated same-sign leptons, $E\!\!\!\!/>$ 50 GeV and $M(\ell^\pm\ell^\pm)>100$ GeV, the number of expected background events is essentially zero \cite{Contino:2008hi,cms}. The number of signal events was found to be 7 corresponding to a significance of $2.6$. The opposite sign two lepton mode may also useful to consider \cite{AguilarSaavedra:2009es}. These various analyses can be combined to enhance the sensitivity to $b'$ while eliminating other new physics explanations of the signal. 

For the case of similar $t'$ and $b'$ masses at or below 600 GeV our results lead us to conclude: 1) a helpful search for $b'\overline{b}'$ production at the LHC can be made via the lepton plus jets mode;  2) the $b'$ mass bumps can be successfully reconstructed by using a $\chi^2$ method; 3) a simple cut on the number of jets is effective at suppressing the $t'$ events (for similar $t'$ and $b'$ masses); 4) a neural network analysis is useful for background discrimination; 5) the resulting significance could be comparable to that of the same-sign leptonic mode. These results are complementary to \cite{atlas,Contino:2008hi,AguilarSaavedra:2009es} and can help experimentalists to decide how to distinguish a $b'$ signal, for example from same-sign leptons, from other new physics.

\section*{Acknowledgments}
We thank anonymous referees for their suggested improvements. This work was supported in part by the Natural Science and Engineering Research Council of Canada. Q.S.~Yan thanks the Theoretical Physics Center for Science Facilities and the Chinese Academy of Sciences in Beijing for hospitality during the final stage of this work. He also thanks Sing-L Cheung for his help.

\end{document}